\documentclass[aps,prd,twocolumn,showpacs,preprintnumbers,amsmath,amssymb,nofootinbib]{revtex4}

\usepackage{graphicx}
\usepackage{dcolumn}
\usepackage{bm}
\usepackage{color}

\begin{document}

\preprint{YITP-10-44}

\title{Analytical model for CMB temperature angular power spectrum 
from cosmic (super-)strings}

\author{Daisuke Yamauchi$^1$}
\email{yamauchi@yukawa.kyoto-u.ac.jp}
\author{Keitaro Takahashi$^2$}
\email{keitaro@a.phys.nagoya-u.ac.jp}
\author{Yuuiti Sendouda$^3$}
\email{sendouda@apc.univ-paris7.fr}
\author{Chul-Moon Yoo$^1$}
\email{yoo@yukawa.kyoto-u.ac.jp}
\author{Misao Sasaki$^1$}
\email{misao@yukawa.kyoto-u.ac.jp}

\affiliation{%
$^1$Yukawa Institute for Theoretical Physics, Kyoto University, Kyoto 606-8502, Japan\\
$^2$Department of Physics and Astrophysics, Nagoya University, Nagoya 494-8602, Japan\\
$^3$APC, Universit\'e Paris 7,
10, rue Alice Domon et L\'eonie Duquet, 75205 Paris cedex 13, France
}%

\date{\today}

\begin{abstract}

We present a new analytical method to calculate 
the small angle CMB temperature angular power spectrum 
due to cosmic (super-)string segments. 
In particular, using our method, we clarify the dependence on 
the intercommuting probability $P$.
We find that the power spectrum is dominated
by Poisson-distributed string segments.
The power spectrum for a general value of $P$
has a plateau on large angular scales and shows
a power-law decrease on small angular scales.
The resulting spectrum in the case of conventional cosmic strings
is in very good agreement with the numerical result obtained by
Fraisse et al..
Then we estimate the upper bound on the dimensionless tension 
of the string for various values of $P$ by assuming that the fraction of 
the CMB power spectrum due to cosmic (super-)strings is
 less than ten percents at various angular scales up to $\ell=2000$.
We find that the amplitude of the spectrum increases as the
intercommuting probability. As a consequence,
strings with smaller intercommuting probabilities are found to be
more tightly constrained.

\end{abstract}

\pacs{Valid PACS appear here}
\maketitle

\section{Introduction}
\label{sec:introduction}

Cosmic strings are line-like topological 
defects formed in the early universe through 
spontaneous symmetry breaking in a wide range of
inflationary models~\cite{Jeannerot:2003qv}.
Since the string tension $\mu$ is directly related to 
the symmetry breaking energy scale, observational
verification of the existence of cosmic strings
will have profound implications to unified theories.
Theoretically, recent developments in string cosmology 
suggest that inflation may be due to motions of 
branes in higher dimensions and various new types
of strings, called cosmic superstrings,
may be formed at the end of
 inflation~\cite{Sarangi:2002yt,Davis:2005dd,Copeland:2009ga,Majumdar:2005qc}.
One of the differences between cosmic superstrings and 
conventional field-theoretic strings
is the value of the intercommuting probability $P$.
It can be significantly smaller than unity for 
cosmic superstrings~\cite{Jackson:2004zg},
while normally it is unity for field-theoretic strings~\cite{Eto:2006db}.

It is known that a moving cosmic (super-)string induces 
a discontinuity in the gravitational potential, 
hence a discontinuity in the cosmic microwave 
background (CMB) temperature map.
This is called the 
Gott-Kaiser-Stebbin (GKS) effect~\cite{Kaiser:1984iv,Gott:1984ef}.
If photons are scattered by a number of moving string segments, 
the observed temperature fluctuations appear as a superposition of
the discontinuities.
The imprint of cosmic strings on CMB has been widely studied.
Using observed CMB anisotropy data, an upper bound on the dimensionless 
energy scale $G\mu$ was 
discussed~\cite{Bevis:2007gh,Pogosian:2003mz,Battye:2010xz,Perivolaropoulos:2005wa}.
The value varies but it is in the range from 
$10^{-7}$ to $10^{-6}$ (for $P=1$).
Although cosmic strings were excluded as a dominant source of 
the observed large angle anisotropy,
a signal due to cosmic strings could still be observed on small 
angular scales with future arcminutes experiments such as 
South Pole Telescope~\cite{Lueker:2009rx} 
and Atacama Cosmology Telescope~\cite{Fowler:2010cy}.
A better theoretical understanding of the temperature fluctuations due to
cosmic (super-)strings may help us distinguish them 
from other secondary effects and may enhance the observability 
of cosmic strings in such experiments. 

Recent numerical 
simulations~\cite{Fraisse:2007nu,Pogosian:2008am,Bevis:2010gj}
show that the small scale CMB temperature angular power spectrum 
due to cosmic strings with $P=1$ behaves as a power law.
One of our purposes of this paper is to derive this power-law
behavior analytically and to extend it to the case of cosmic
strings with $P<1$.

In \cite{Takahashi:2008ui,Yamauchi:2010vy}, 
we computed the one-point probability distribution function (pdf)
with a simple model of long curved string segments and kinks.
It was found that the one-point pdf is dominated by
a Gaussian component due to frequent scatterings by long
straight segments, together with small non-Gaussian tails due to
close encounters with kinks and a small asymmetry,
namely skewness, from the correlation between curvatures 
and velocities of string segments.
Therefore, as far as the power spectrum is concerned,
it is sufficient to consider only the contribution from long straight 
string segments.

In this paper, we present a new analytical method to calculate 
the CMB temperature angular power spectrum due to cosmic (super-)string segments.
Our formalism is similar to the halo formalism for the Sunyaev-Zel'dovich 
effect~\cite{Komatsu:2002wc,Komatsu:1999ev,Cole:1989vx}.
We adopt a simple model of string network for general values
of the intercommuting probability $P$.
We consider long straight string segments which are 
located randomly between the last scattering surface (LSS) 
and the present time consistently with the string network model.
We find that the angular power spectrum
is dominated by Poisson-distributed segments.
Then we find it is possible to derive the dependence 
of the CMB spectrum on the intercommuting probability $P$
explicitly.

This paper is organized as follows.
In section \ref{sec:VOS}, we give basic equations governing
a string network which incorporate the intercommuting probability $P$.
In section \ref{sec:GKS}, we give an explicit form of the GKS effect
and perform the Fourier transformation analytically.
In section~\ref{sec:segment approach}
we introduce a formalism called `the segment formalism' 
and derive an analytical formula for the angular power spectrum
due to cosmic (super-)strings.
In section \ref{sec:constraint}, we calculate an upper bound
on the dimensionless tension $G\mu$ under appropriate
assumptions.
Finally, we summarize our results in \ref{sec:summary}.

\section{String network model}
\label{sec:VOS}

We first give basic equations governing string 
network incorporating intercommuting probability $P$, 
following~\cite{Takahashi:2008ui,Yamauchi:2010vy}.
A string worldsheet can be 
described by $x^\mu=x^\mu(\sigma^a)$, 
where $x^\mu$ and $\sigma^a$ are the spacetime coordinates
and the worldsheet coordinates, respectively. 
Let us consider string dynamics in a
Friedmann-Lema\^itre-Robertson-Walker 
universe with the metric 
\begin{eqnarray}
ds^2=a^2(\eta )\left( -d\eta^{2} +d{\bm r}^2\right)\,. 
\end{eqnarray}
We choose the temporal gauge : $\sigma^{0}=\eta\,,\ \sigma^{1}=\sigma\,,\ 
\dot{\bm r}\cdot{\bm r}^{\prime}=0\,,$
where the bold letters denote $3$-vectors on the comoving space
and the dot and the prime denote the derivatives
with respect to $\eta$ and $\sigma$, respectively.

In the velocity-dependent one-scale model,
a string segment has two properties, the length $\xi$ and 
the root-mean-square velocity $v_{\rm rms}$.
The typical length $\xi$ is defined by $\xi\equiv\sqrt{\mu /\rho_{\rm seg}}$
where $\rho_{\rm seg}$ is the total string energy density.
In our treatment, we also take account of
the energy loss due to loop formation.
The characteristic time scale for loop formation is $\sim\xi /(Pv_{\rm rms})$ and 
the energy loss can be described as $\sim \tilde{c}Pv_{\rm rms}\rho_{\rm seg}/\xi$
where we have introduced $\tilde{c}$ as a constant which represents 
the efficiency of loop formation.
Assuming $a(t)\propto t^{\beta}$ with the physical time $t=\int a(\eta )d\eta$, 
the equations of motion for $\gamma$ and $v_{\rm rms}$ are 
given by~\cite{Takahashi:2008ui,Yamauchi:2010vy}
\begin{eqnarray}
&&\frac{t}{\gamma}\frac{d\gamma}{dt}
=1-\beta -\frac{1}{2}\beta\tilde{c}Pv_{\rm rms}\gamma
-\beta v_{\rm rms}^2
\,,\label{eq:gamma eq}\\
&&\frac{dv_{\rm rms}}{dt}
=(1-v_{\rm rms}^2)H
\Bigl[ k(v_{\rm rms})\gamma -2v_{\rm rms}\Bigr]\,, 
\label{eq:v_rms eq}
\end{eqnarray}
where $k(v_{\rm rms})=(2\sqrt{2}/\pi )(1-8v_{\rm rms}^6)/(1+8v_{\rm rms}^6)$~\cite{Martins:2000cs}.
Hereafter we assume a matter-dominated era, $\beta =2/3$,
and we use $\tilde{c}\approx 0.23$ as the 
standard value~\cite{Martins:2003vd}.

It is known that a string network approaches the so-called 
scaling regime where the characteristic scale grows 
with the Hubble horizon size~\cite{Kibble:1984hp,Ringeval:2005kr}. 
We assume that the scaling is already realized
by the time of the last scattering surface (LSS) 
and this means that $\gamma ,v_{\rm rms}$ are constant in time.
For small $\tilde{c}P$, we can solve 
Eqs.~\eqref{eq:gamma eq}, \eqref{eq:v_rms eq} approximately
as~\cite{Takahashi:2008ui}
\begin{eqnarray}
v_{\rm rms}^2\approx\frac{1}{2}\biggl[
1-\frac{\pi}{3\gamma}\biggr]\,,
\quad 
\gamma=\gamma (P)\approx\sqrt{\frac{\pi\sqrt{2}}{3\tilde{c}P}}
\,.\label{eq:scaling sol}
\end{eqnarray}

\section{Temperature fluctuations due to a string segment}
\label{sec:GKS}

In order to discuss temperature fluctuations on small angular scales,
we focus on a small patch of sky, and
consider a straight string segment at
the position ${\bm r}(\eta ,\sigma )$ where $\eta$ and $\sigma$
are the time and position on the string worldsheet. 
We introduce a vector ${\bm X}(\sigma )$ describing
the comoving position of an observer relative to that of the string:
\begin{eqnarray}
{\bm X}(\sigma )={\bm r}_{\rm obs}
-{\bm r}(\sigma ,\eta_{\rm lc}(\sigma ))\,,
\end{eqnarray}
where $\eta_{\rm lc}(\sigma )$ is the conformal time along the intersection
of the observer's past light-cone and the string worldsheet,
$\eta_{\rm obs}-\eta_{\rm lc}(\sigma )=|{\bm X}(\sigma )|$.
On small scales, the temperature fluctuation due to a string segment
in the direction ${\bm n}$, which is the unit vector along the line-of-sight,
is given by~\cite{Stebbins:1994ng,Stebbins:1987va,Hindmarsh:1993pu}
\begin{eqnarray}
&&\frac{\Delta T}{T}=-4G\mu\int_{\Sigma} d\sigma
\frac{{\bm X} ^{\perp}\cdot{\bm u}}{|{\bm X}^{\perp}|^2}
\,,\label{eq:temp fluc in small angle}
\end{eqnarray}
where ${\bm X}^{\perp}(\sigma )
={\bm X}(\sigma )-({\bm X}(\sigma )\cdot{\bm n}){\bm n}$,
and ${\bm u}$ is defined by
\begin{eqnarray}
{\bm u}(\sigma )\equiv\biggl[
\dot{\bm r}
-\left(\frac{{\bm n}\cdot{\bm r}'}
{1+{\bm n}\cdot\dot{\bm r}}\right)
{\bm r}'\biggr]_{\eta =\eta_{\rm lc}(\sigma )}\,,
\end{eqnarray}
and $\Sigma$ denotes the intersection of the observer's past light-cone
with the string worldsheet, along which the integration is to be performed.
We have adopted the small angle approximation
and neglected higher order terms since their contributions
are very small in general~\cite{Yamauchi:2010vy}.

Hereafter, we assume that the string segment is exactly straight and 
uniformly moving, that is
${\bm r}'={\rm const.}\,,\dot{\bm r}={\rm const.}$.
For an exactly straight and uniformly moving segment,
Eq.~\eqref{eq:temp fluc in small angle}
reduces to~\cite{Stebbins:1994ng,Yamauchi:2010vy}
\begin{eqnarray}
\frac{\Delta T}{T}
=4G\mu\frac{|\dot{\bm r}|}{\sqrt{1-\dot{\bm r}^2}}\alpha_{\rm seg}
\int_{\Sigma} d\sigma
\frac{({\bm n}\times{\bm X}^{\perp})\cdot\frac{d{\bm X}^{\perp}}{d\sigma}}
{|{\bm X}^{\perp}|^2}\,,
\end{eqnarray}
where we have introduced 
\begin{eqnarray}
\alpha_{\rm seg}={\bm n}\cdot
\frac{{\bm r}'}{|{\bm r}'|}\times
\frac{\dot{\bm r}}{|\dot{\bm r}|}
\,.
\label{eq:alpha_seg}
\end{eqnarray}
The position of the exactly straight and uniformly 
moving segment can be written as
\begin{eqnarray}
a{\bm X}^{\perp}(\sigma )=d_A{\bm \vartheta}
+a\sigma\biggl|\frac{d{\bm X}^{\perp}}{d\sigma}\biggl|\,{\bm e}\,,
\end{eqnarray}
where $d_A$ denotes the angular diameter distance from the
observer, 
${\bm e}\propto d{\bm X}^{\perp}/d\sigma=\mbox{const.}$
is the unit vector along the string,
and ${\bm \vartheta}$ is the angular position vector
relative to the middle point of the 
segment in a small patch of sky.

Here, let us introduce a set of orthonormal basis 
vectors $({\bm s}_1,{\bm s}_2,{\bm n})$, where the vectors
$({\bm s}_1,{\bm s}_2)$ span a patch of sky orthogonal to
the line of sight.
Then ${\bm e}$ and ${\bm \vartheta}$ may be expressed as
\begin{eqnarray}
{\bm e}&=&\cos\varphi_{e}{\bm s}_1+\sin\varphi_{e}{\bm s}_2\,,
\cr
\cr
{\bm \vartheta}
&=&\vartheta\left(\cos\varphi{\bm s}_1+\sin\varphi{\bm s}_2\right)\,,
\end{eqnarray}
where the ranges of the coordinates are
$0\leq \vartheta<\infty$ and $0\leq\varphi\leq 2\pi$
in the flat sky approximation (which is valid under the
small angle approximation).

We also need explicit expressions of the components
of the vectors ${\bm r}'$ and $\dot{\bm r}$,
which determine the configuration of a string,
where we adopt the gauge condition 
${\bm r}'\cdot\dot{\bm r}=0$.
Thus the number of independent degrees of freedom 
is $6-1=5$. A convenient parametrization is~\cite{Yamauchi:2010vy}
\begin{eqnarray}
&&{\bm r}'=|{\bm r}'|\Bigl[
\sin\theta\cos\phi{\bm s}_{1}
+\sin\theta\sin\phi{\bm s}_{2}
+\cos\theta{\bm n}
\Bigr]\,,
\nonumber\\
&&\dot{\bm r}=|\dot{\bm r}|\Bigl[
(-\sin\psi\cos\theta\cos\phi
-\cos\psi\sin\phi ){\bm s}_{1}
\nonumber\\
&&\ \ \ \ \ \ \ \ \ 
+(-\sin\psi\cos\theta\sin\phi
+\cos\psi\cos\phi ){\bm s}_{2}
\nonumber\\
&&\ \ \ \ \ \ \ \ \ 
+\sin\psi\sin\theta{\bm n}
\Bigr]\,,
\end{eqnarray}
where the ranges of the angular parameters are
$0\leq \theta\leq \pi$, $0\leq\phi\leq 2\pi$,
$0\leq\psi\leq 2\pi$.

For notational simplicity, we introduce 
${\bm \Theta}\equiv\{\theta ,\phi ,\psi\}$ to
denote the set of the angular parameters.
Then, Eq.~\eqref{eq:alpha_seg} reduces to
\begin{eqnarray}
\alpha_{\rm seg}({\bm \Theta})=\cos\psi\sin\theta\,.
\end{eqnarray}
Note that the unit tangent vector ${\bm e}$ can be written 
as a function of the parameter ${\bm \Theta}$, 
$\varphi_{e}=\varphi_{e}({\bm \Theta})$. However,
as we see below, it turns out that we do not need
an explicit expression for $\varphi_{e}$.

In our calculation, we consider only a segment 
of a long straight string with length $\sim\xi$ at each scattering.
Therefore we take the range of integration over $\sigma$
as $a|d{\bm X}^{\perp}/d\sigma|\,|\sigma |\leq\xi$.
Then the temperature deviation 
is~\cite{Kaiser:1984iv,Gott:1984ef,Yamauchi:2010vy}
\begin{eqnarray}
&&\frac{\Delta T}{T} ({\bm \vartheta})
=-4G\mu\frac{|\dot{\bm r}|}{\sqrt{1-\dot{\bm r}^2}}\alpha_{\rm seg}
\nonumber\\
&&\times\biggl\{
\arctan\left[\frac{\ell_{\rm co}^{-1}+\vartheta\cos\hat\varphi}{\vartheta\sin\hat\varphi}\right]
+\arctan\left[\frac{\ell_{\rm co}^{-1}-\vartheta\cos\hat\varphi}{\vartheta\sin\hat\varphi}\right]
\Biggr\}
\,,\nonumber\\
\end{eqnarray}
where $\hat\varphi=\varphi -\varphi_{e}$ and we have introduced
the angular scale $\ell_{\rm co}$
corresponding to the correlation length of the segment,
\begin{eqnarray}
\ell_{\rm co}\equiv\frac{d_A}{\xi}=\gamma Hd_A\,.
\end{eqnarray}

As mentioned above, since we focus on small angular scales, 
the flat sky approximation is valid.
Therefore we can perform the Fourier transformation of the 
temperature fluctuation analytically.
The Fourier transform is 
defined by~\cite{Liddle:2000cg,White:1997wq}
\begin{eqnarray}
&&a_{\bm \ell}
=\frac{1}{2\pi}\int d^2{\bm \vartheta}
\frac{\Delta T}{T} ({\bm \vartheta})
e^{-i{\bm \ell}\cdot{\bm \vartheta}}\,.
\end{eqnarray}
Then we obtain
\begin{eqnarray}
a_{\bm \ell}
=-\frac{8iG\mu |\dot{\bm r}|\alpha_{\rm seg}({\bm\Theta})}
{\sqrt{1-\dot{\bm r}^2}\ell^{2}}
\tan\hat\varphi_{\ell}
\sin\left(\frac{\ell}{\ell_{\rm co}}\cos\hat\varphi_{\ell}\right),
\label{eq:FT of GKS}
\end{eqnarray}
where ${\bm \ell}=\ell(\cos\varphi_{\ell}{\bm s}_1+\sin\varphi_{\ell}{\bm s}_2)$
 and $\hat\varphi_{\ell}=\varphi_{\ell}-\varphi_{e}$.
As apparent from the above expression, the result does not
depend explicitly on $\varphi_e$ but only through
the angle relative to the string segment, 
$\hat\varphi_{\ell}=\varphi_{\ell}-\varphi_{e}$. This is the reason
why it is unnecessary to express $\varphi_e$ explicitly in terms
of the angular parameters ${\bm\Theta}$.

\section{Segment formalism}
\label{sec:segment approach}

In order to compute the angular power spectrum of
the temperature fluctuations due to cosmic (super-)strings,
we use what we call the {\it segment formalism}, by adapting from
the halo formalism for the Sunyaev-Zel'dovich 
effect~\cite{Komatsu:2002wc,Komatsu:1999ev,Cole:1989vx}.
Since the observed sky map of temperature fluctuations due to segments 
appears as a superposition of those due to each segment,
the Fourier transform of the total temperature fluctuations,
$a_{\bm \ell}^{\rm tot}$,
can be decomposed into each contribution of each string segment.
In our treatment, we first introduce a segment index ``$i$''
to denote the contribution from each segment
between LSS and the present.
Then we have
\begin{eqnarray}
&&a_{\bm \ell}^{\rm tot}(\{{\bm \Theta}_i,z_i\} )
=\sum_{i=1}^{N}a_{\bm \ell}
({\bm \Theta}_i,z_i)
\equiv\sum_{i=1}^{N}a_{\bm \ell}^{(i)}
\,,
\end{eqnarray}
where ${\bm \Theta}_i$ and $z_i$ are the segment configuration parameters
and the redshift, respectively, of the $i$-th string segment.
$N$ is the total number of the string segments.

Asuming the statistical isotropy of the CMB,
the angular power spectrum can be written as
\begin{eqnarray}
&&C_{\ell}=
\int\frac{d\varphi_{\ell}}{2\pi}
\Big\langle a_{\bm \ell}^{\rm tot}(\{{\bm \Theta}_i,z_i\} )
a_{\bm \ell}^{\rm tot}(\{{\bm \Theta}_j,z_j\} )^*\Big\rangle
\nonumber\\
&&\ \ \ \ 
=\int\frac{d\varphi_{\ell}}{2\pi}\Big\langle\sum_{i} 
\bigl| a_{\bm \ell}^{(i)}\bigl|^{2}
\Big\rangle
+\int\frac{d\varphi_{\ell}}{2\pi}\Big\langle\sum_{i\neq j} 
a_{\bm \ell}^{(i)}
{a_{\bm \ell}^{(j)}}^*\Big\rangle
\nonumber\\
&&\ \ \ \ 
\equiv C_{\ell}^{1\text{seg}}+C_{\ell}^{2\text{seg}}
\,,
\end{eqnarray}
where the integral over $\varphi_\ell$ is the large $\ell$ approximation
of the sum over the azimuthal eigenvalues $m$ ($-\ell\leq m\leq\ell$),
and $\langle\cdots\rangle$ denotes the ensemble average.
The ensemble average can be calculated 
by averaging over the parameter space,
\begin{eqnarray}
\Big\langle\cdots\Big\rangle\rightarrow 
\prod_{i}\biggl[
\frac{1}{N}\int dz_{i}\frac{dV}{dz_{i}}
\int d{\bm \Theta}_i\cdot\frac{dn}{d{\bm \Theta}_i}
\biggr]\cdots\,,
\end{eqnarray}  
where $(dV/dz)dz$ is the differential comoving volume element
at redshift $z$, $(dn/d{\bm\Theta})\cdot d{\bm \Theta}$ 
is the comoving number density of 
string segments with the parameters in the range 
$[{\bm \Theta},{\bm \Theta}+d{\bm \Theta}]$.
Note that the total number of the segments can be rewritten as
$N=\int dz(dV/dz)\int (dn/d{\bm \Theta})\cdot d{\bm \Theta}$.
We assume uniform distributions $P(\cos\theta )=1/2$, $P(\phi)=P(\psi)=1/2\pi$
of the parameters. Then, the number density of the segments can be
estimated as
\begin{eqnarray}
&&\frac{dn}{d{\bm \Theta}_i}\cdot d{\bm \Theta}_i
\approx H^3\gamma^{3}
\frac{d(\cos\theta_{i} )d\phi_{i} d\psi_{i}}{2(2\pi )^2}
\,,
\end{eqnarray}
where we have adopted the scaling ansatz and
used Eq.~\eqref{eq:scaling sol}.

\begin{figure}[tbp]
 \begin{center}
  \includegraphics[width=80mm]{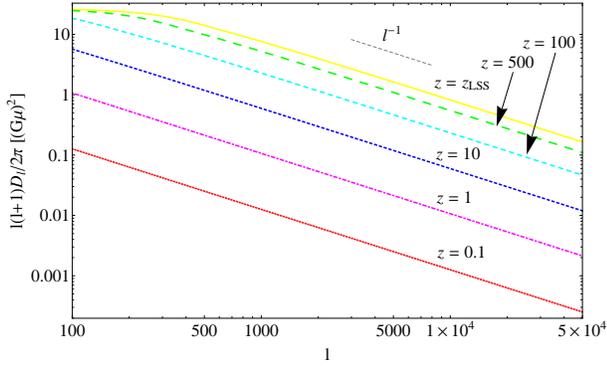}
 \end{center}
 \caption{The contribution to $C_\ell$
from each logarithmic interval of $1+z$, $\ell(\ell+1)D_\ell(z)/(2\pi)$,
in the case of $P=1$ in units of $(G\mu )^2$.
Here it is shown as a function of $\ell$ for various values of $z$.
From top to bottom, $z=z_{\rm LSS}$, $500$, $100$, $10$, $1$ and $0.1$.}
 \label{fig:Pw_P=1_z}
\end{figure}

As mentioned in Sec.~\ref{sec:introduction}, 
long straight string segments are assumed to be
distributed randomly between LSS and the present 
consistently with the string network model.
This implies there is no correlation between two segments, 
$\langle a_{\bm \ell}^{(i)}a_{\bm \ell}^{(j)}\rangle =0$
for $i\neq j$.
If we consider a more general string network,
there may be some nonzero contribution 
from the segment-segment correlation.
However, the segment-segment correlation at a redshift $z$
would be dominated by the contribution from 
$\ell \approx\ell_{\rm co}(z)$
and the smallest scale at which the segment-segment correlation
could be important is determined by $\ell_{\rm co}$ at $z=z_{\rm LSS}$,
\begin{eqnarray}
\ell_{\rm co}(z)\leq \ell_{\rm co}(z_{\rm LSS})
\approx 156\left(\frac{\tilde{c}P}{0.23}\right)^{-1/2}\,,
\label{eq:ell_co}
\end{eqnarray}
where we have put $z_{\rm LSS}\approx 1100$.
Therefore, 
the angular power spectrum on small scales,
$\ell >\ell_{\rm co}(z_{\rm LSS})$, 
will be dominated formally by the contribution
of the sum of $N$ statistically independent segments,
even if the segment-segment correlation is taken into account.
That is,
\begin{eqnarray}
&&C_{\ell}\approx\int^{z_{\rm LSS}}_0dz\frac{dV}{dz}
\int d{\bm \Theta}\cdot\frac{dn}{d{\bm \Theta}}\ 
{\cal G}_{\ell}({\bm \Theta},z)
\,,
\end{eqnarray}
with
\begin{eqnarray}
&&{\cal G}_{\ell}({\bm \Theta},z)
=\int\frac{d\hat\varphi_{\ell}}{2\pi}
\bigl| a_{\bm \ell}({\bm \Theta},z)\bigl|^{2}
\,,
\end{eqnarray}
where the integral over $\varphi_\ell$
has been replaced by that over 
$\varphi_{\ell}\rightarrow \hat\varphi_{\ell}=\varphi_{\ell}-\varphi_{e}$
without loss of generality.

\begin{figure}[tbp]
 \begin{center}
  \includegraphics[width=80mm]{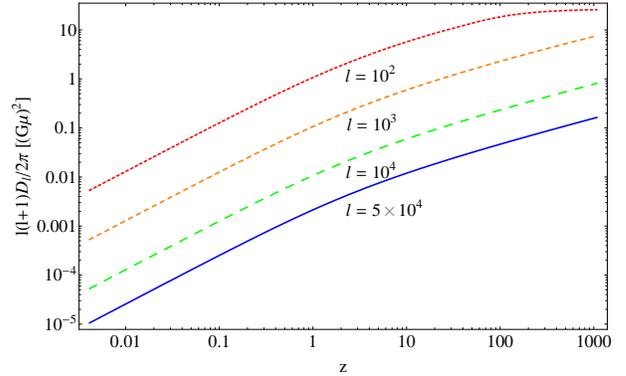}
 \end{center}
 \caption{The same as Fig.~\ref{fig:Pw_P=1_z} but
as a function of $z$ for various values of $\ell$.
From top to bottom, $\ell =10^2$, $10^3$, $10^4$ and $5\times 10^4$.}
 \label{fig:Pw_P=1_l}
\end{figure}

Assuming that $|\dot{\bm r}|=v_{\rm rms}$ and that 
the universe is matter-dominated, we have
$\ell_{\rm co}(z)\approx 2\gamma (\sqrt{1+z}-1)$
and $dV/dz=16\pi H^{-3}(1-(1+z)^{-1/2})^2$.
Then we obtain an explicit form of $C_\ell$ as
\begin{eqnarray}
\label{eq:formula}
C_{\ell}&=&\int_{0}^{z_{\rm LSS}}\frac{dz}{1+z}D_\ell (z)\,,
\end{eqnarray}
where
\begin{eqnarray}
D_{\ell}(z)&=&(1+z)\frac{dV}{dz}\int d{\bm \Theta}\cdot\frac{dn}{d{\bm \Theta}}
\ {\cal G}_{\ell}({\bm \Theta},z)
\cr
&\approx&
\frac{2\pi (8\gamma )^{3} v_{\rm rms}^2(G\mu )^2}{3(1-v_{\rm rms}^2)\ell^{4}}
\left( \sqrt{1+z}-1\right)^{2}
\nonumber\\
&\times&
\int^{\pi}_{-\pi}\frac{d\hat\varphi_{\ell}}{2\pi}\tan^{2}\hat\varphi_{\ell}
\sin^{2}\left(\frac{\ell}{\ell_{\rm co}(z)}\cos\hat\varphi_{\ell}\right).
\end{eqnarray}
Here $D_{\ell}(z)$ is the redshift distribution of $C_{\ell}$,
which tells us which redshift $z$ contributes most for a given $\ell$,
or which $\ell$ contributes most at a given $z$.
In Figs.~\ref{fig:Pw_P=1_z} and \ref{fig:Pw_P=1_l},
we plot $D_{\ell}(z)$ for $P=1$.
We see that the large $z$ contribution dominates for a given $\ell$.
This may be explained by the fact that 
the number of segments per unit redshift bin
becomes large as $z$ increases.
In other words, the contribution from small $z$ is negligible,
justifying the use of the small angle approximation
in Eq.~(\ref{eq:temp fluc in small angle}).

\begin{figure}[tbp]
 \begin{center}
  \includegraphics[width=80mm]{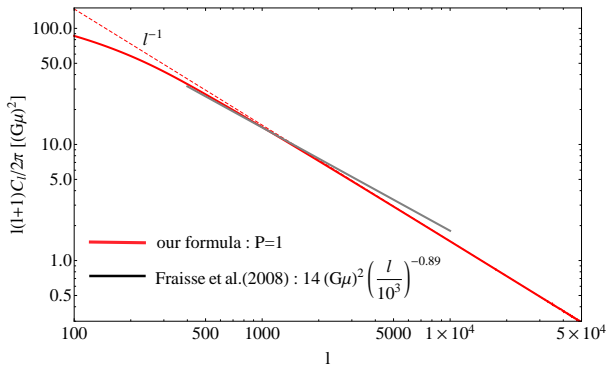}
 \end{center}
 \caption{The angular power spectrum for $P=1$ in units of $(G\mu )^2$.
The red solid line is our result given by Eq.~\eqref{eq:formula}.
The red dotted line shows the behavior $\propto \ell^{-1}$.
The gray solid lines are power-law fit to previous numerical result
by Fraisse et al.~\cite{Fraisse:2007nu}.
}
 \label{fig:Pw_total_P=1}
\end{figure}

\begin{figure}[tbp]
 \begin{center}
  \includegraphics[width=80mm]{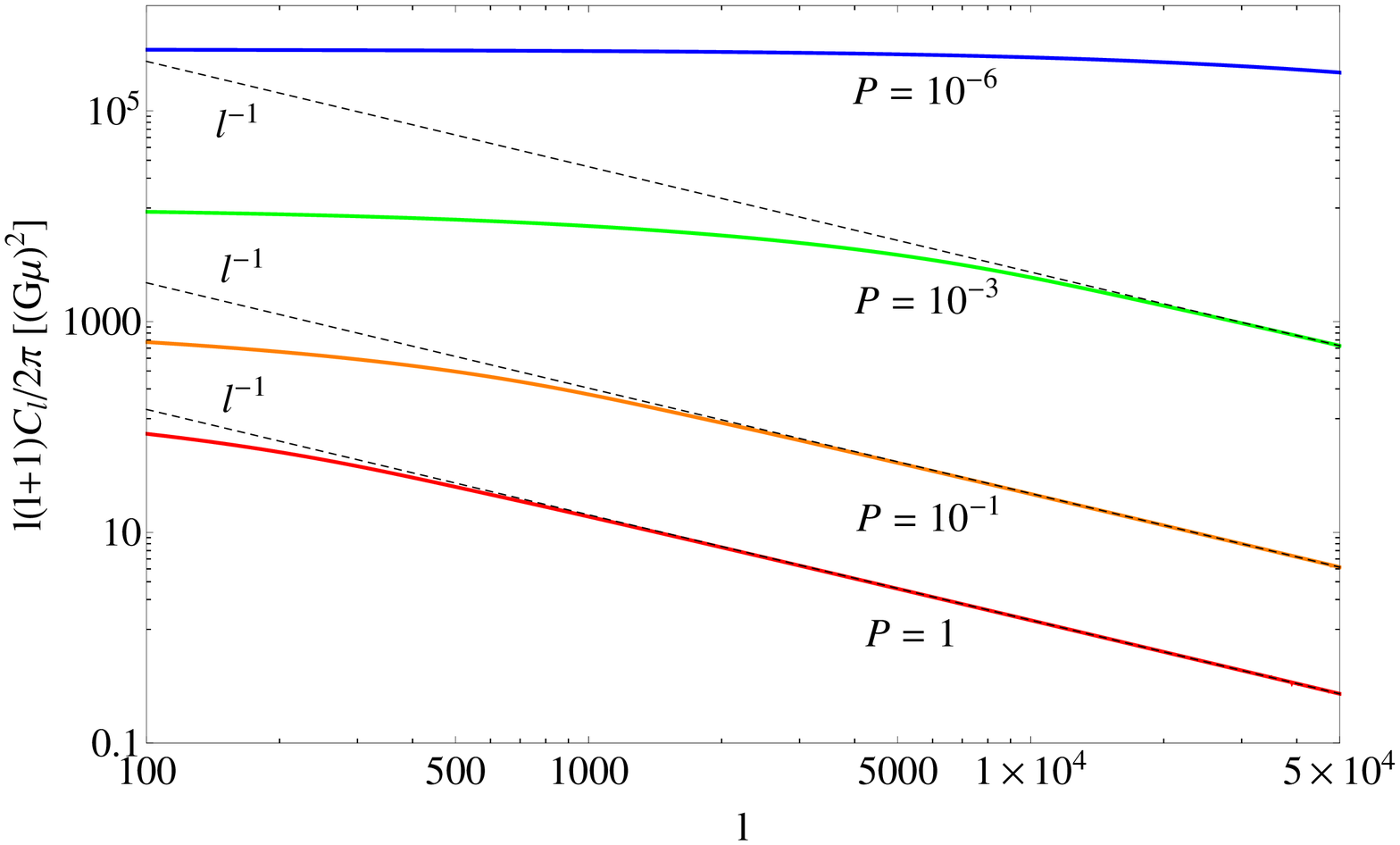}
 \end{center}
 \caption{The angular power spectrum given by Eq.~\eqref{eq:formula}
in units of $(G\mu )^2$.
The curves are, from bottom to top, for $P=1$ (red), $P=10^{-1}$ (orange),
$P=10^{-3}$ (green), and $P=10^{-6}$ (blue).
}
 \label{fig:Pw_total_P}
\end{figure}

The total CMB temperature angular power spectrum 
in the case of $P=1$ is shown in Fig.~\ref{fig:Pw_total_P=1}.
As seen from it, a typical amplitude of the power spectrum
at $\ell =10^3$ is
$[\ell (\ell +1)/2\pi ]C(\ell =10^{3})\approx 14(G\mu )^2$,
and it behaves as $\ell^{-1}$ for large 
$\ell (\gg\ell_{\rm co}(z_{\rm LSS}))$ 
(see also \cite{Hindmarsh:2009qk,Hindmarsh:1993pu,Regan:2009hv}) while
it has a plateau for small $\ell (\lesssim\ell_{\rm co}(z_{\rm LSS}))$.
For comparison, we also plot power-law fit to previous numerical result:
\begin{eqnarray*}
&&\frac{\ell (\ell +1)}{2\pi}C_{\ell}
\approx 14(G\mu )^2\left(\frac{\ell}{1000}\right)^{-0.89}
\cr
&&\ \ \ \ \ 
\mbox{for}~400\leq \ell\leq 10^4
~
\mbox{(Fraisse et al.~\cite{Fraisse:2007nu})}\,.
\end{eqnarray*}
As clearly seen, our result agrees very well with the numerical 
result by Fraisse et al.~\cite{Fraisse:2007nu}.
This strongly supports the validity of our approach.
We should note that the spectrum obtained here as
well as the one obtained by Fraisse et al.~\cite{Fraisse:2007nu} ignores
fluctuations induced by strings on the last scattering surface (LSS).
On small angular scales, the LSS contributions are damped and
the GKS contributions dominate. 
Nevertheless the LSS contributions are not completely negligible even 
at $\ell\approx 4000$, which seems to be the case in \cite{Bevis:2010gj,Pogosian:2008am}.
Thus, one cannot really directly compare the spectra 
from Bevis et al.~\cite{Bevis:2010gj} and 
Pogosian et al.~\cite{Pogosian:2008am} using the code CMBACT~\cite{Pogosian:1999np} 
to those of our result and Fraisse et al.~\cite{Fraisse:2007nu} 
for $\ell <3000$~\cite{Pogosian:private}.
In \cite{Bevis:2010gj}, the authors showed that the $\ell^{-2}$ behavior
for $1000<\ell <3000$ becomes much closer to $\ell^{-1}$
when only the string sources after recombination are taken into account.
Therefore, our result agrees not only with \cite{Fraisse:2007nu} but also with
\cite{Bevis:2010gj} as far as only the GKS component is considered.
\footnote{Our result does not agree well
with Pogosian et al.~\cite{Pogosian:2008am}.
They used CMBACT~\cite{Pogosian:1999np}
and obtained a larger amplitude for $100<\ell <3000$
and an approximate $\ell^{-1.5}$ decay at high $\ell$.
The difference in the amplitude is probably due to different scaling parameters
such as $\gamma$ and $v_{\rm rms}$ between theirs and ours.}
\\

In order to investigate the dependence on the intercommuting probability $P$,
the angular power spectrum~\eqref{eq:formula} is computed for 
various $P$. The results are shown in Fig.~\ref{fig:Pw_total_P}.
We see that the overall amplitude of the spectrum increases as $P$
decreases. This is because of the factor $\gamma^{3}\propto P^{-3/2}$
in the formula~\eqref{eq:formula}, which describes the fact that
the density of cosmic string segments is larger for smaller $P$.
Also since $\ell_{\rm co}(z_{\rm LSS})\propto P^{-1/2}$,
we see that the transition from the plateau
to the power law occurs at larger $\ell$ for smaller $P$.
These properties of the power spectrum may become a useful
tool to distinguish the value of $P$ in future experiments.

\section{Constraints on string tension}
\label{sec:constraint}

\begin{figure}[tbp]
 \begin{center}
  \includegraphics[width=80mm]{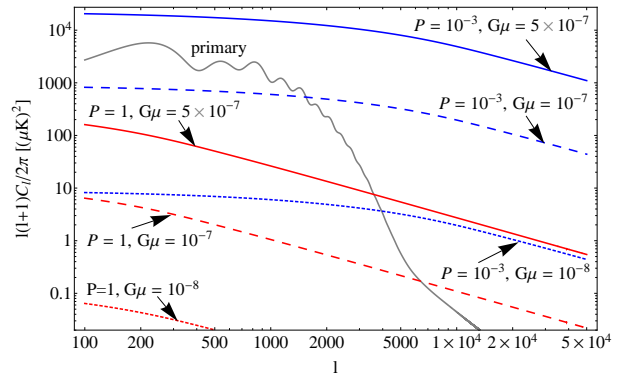}
 \end{center}
 \caption{The angular power spectrum in units of $\mu$K$^2$.
The red lines are for $P=1$ and the blue lines are $P=10^{-3}$.
For both cases $G\mu =5\times10^{-7}$, $10^{-7}$ and $10^{-8}$
from top to bottom.
For comparison, the primary spectrum is shown in gray.}
 \label{fig:power_prim_vs_CS}
\end{figure}

Let us discuss possible constraints on the string tension
from our result. We plot the angular power spectrum
for various values of $P$ and $G\mu$ in Fig.~\ref{fig:power_prim_vs_CS}.
For comparison, we also plot the primary spectrum.
An interesting observation is that as $P$ decreases 
the amplitude due to strings increases, hence the tension
of strings with smaller $P$ is more tightly constrained.

\begin{figure}[tbp]
 \begin{center}
  \includegraphics[width=80mm]{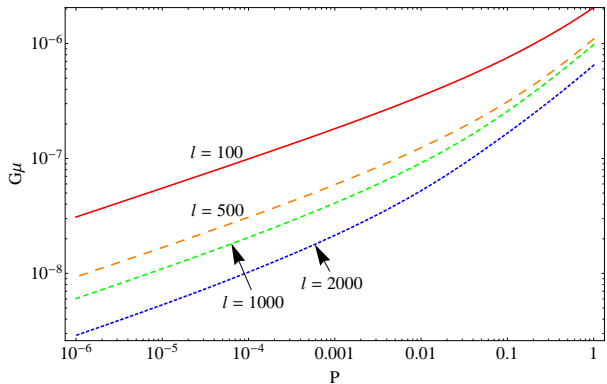}
 \end{center}
 \caption{Upper bound on $G\mu$ as a function of $P$
at $\ell =10^2$, $5\times 10^2$, $10^{3}$ and $2\times 10^3$,
assuming that the fraction of the spectrum due to
cosmic (super-)strings is less than $10\%$.}
 \label{fig:P_vs_Gmu_bound}
\end{figure}

It was pointed out in \cite{Bevis:2007gh} that the CMB anisotropy
spectrum is consistent with the presence of cosmic strings
if the fraction of the power spectrum due to cosmic strings 
is about $10\%$ or less at $\ell =10$.
In this paper we adopt a similar criterion and
drive an upper bound on $G\mu$ as a function of $P$.
Specifically, we consider the condition that the fraction 
of the power spectrum due to cosmic (super-)strings is less than 
$10\%$ at $\ell =10^2$, $5\times 10^2$, $10^{3}$ and
$2\times 10^3$. The result is shown in Fig.~\ref{fig:P_vs_Gmu_bound}.
As expected, the upper bound on $G\mu$ decreases as $P$ decreases,
because the amplitude of the power spectrum increases.
Also, we see that the constraint becomes severer for larger $\ell$
because the contribution from cosmic strings decays very slowly as
$\ell$ increases unlike the case of the primordial anisotropy which
shows exponential dumping.
For example, the upper bound at 
$\ell=10^2$ is $2.1\times 10^6$ for $P=1$ and $3.1\times 10^{-8}$ for $P=10^{-6}$,
while that at $\ell =10^3$ is
$9.8\times 10^{-7}$ for $P=1$ and $6.0\times 10^{-9}$ for $P=10^{-6}$.

\section{Summary}
\label{sec:summary}

In this paper, we presented a new analytical method to calculate 
the small angle CMB power spectrum due to cosmic (super-)string
segments, and investigated the dependence of
the power spectrum on the intercommuting probability $P$.

We found that the angular power spectrum on small scales
can be well approximated by the GKS effect
due to Poisson-distributed mutually independent segments.
Then we derived an analytical formula for the power spectrum
valid for general values of $P$.
The obtained power spectrum for $P=1$, that is for
conventional cosmic strings, was found to agree
very well with the numerical result obtained by
Fraisse et al.~\cite{Fraisse:2007nu}.
This strongly supports the validity of our approach,
hence allows us to discuss the dependence on $P$
with high confidence.

The angular power spectrum is found to behave as $\ell^{-1}$ for 
large $\ell (\gg\ell_{\rm co}(z_{\rm LSS}))$ 
and have a plateau for small $\ell (\lesssim\ell_{\rm co}(z_{\rm LSS}))$,
where $\ell_{\rm co}(z_{\rm LSS})$ is the angular scale corresponding
 the correlation length at LSS. Since $\ell_{\rm co}(z_{\rm LSS})$ is
proportional to $P^{-1/2}$ in the scaling regime, 
the transition from a plateau to the power-law behavior 
is found to occur at larger $\ell$ as $P$ decreases.
We should note, however, that the plateau region of the power spectrum
may have additional modifications because 
the segment-segment correlatoin may not be negligible on
scales $\ell <\ell_{\rm co}(z_{\rm LSS})$, though its effect
on the spectrum is expected to be small if not negligible.

Then using our result, we discussed an upper bound on 
the dimensionless tension $G\mu$ as a function of $P$.
We assumed that the fraction of the CMB spectrum due to
cosmic (super-)strings is less than $10\%$,
and derived an upper bound at $\ell =10^2$, $5\times 10^2$, $10^{3}$
and $2\times 10^3$.
We found that strings with small $P$ are more tightly constrained.
This can be naturally explained by the fact that 
the amplitude of the spectrum increases as $P$ decreases
because of the increase in the number density of strings.

These properties of the power spectrum are
distinguishable features of cosmic superstrings that
generally have a small intercommuting probability $P$. 
They may be used to detect cosmic superstrings 
in future experiments.

Finally, we comment on the bispectrum due to string segments.
It is easy to see that in the present approach we have
$\big\langle (\alpha_{\rm seg})^{2m+1}\big\rangle =0$ and
$\big\langle (\alpha_{\rm seg})^{2m}\big\rangle\neq 0$ for $m=0,1,\cdots$.
This implies a vanishing bispectrum.
To obtain a non-vanishing bispectrum it is necessary to 
take the correlation between the velocity and the curvature 
of a string segment into account~\cite{Hindmarsh:2009qk,Regan:2009hv}
(see also \cite{Takahashi:2008ui,Yamauchi:2010vy}
for discussion on the skewness of a one-point probability 
distribution function).
It is left for future work to include such correlations 
in our segment formalism, and
calculate the non-Gaussianity of the spectrum (e.g.~\cite{Zhang:2007psa}).
Since the non-Gaussian features are expected to significantly depend on 
the intercomuting probability, they may be used to distinguish 
cosmic superstrings from conventional field theoretic cosmic strings.

\section*{Acknowledgments}

We thank M.~Hindmarsh, L.~Pogosian, D.~A.~Steer and S.~-H.~H.~Tye
for valuable comments and useful suggestions.
We also thank the organizers and participants of 
workshop : ``The non-Gaussian Universe'' (YITP-T-09-05) 
and long-term workshop : ``Gravity and Cosmology 2010'' (YITP-T-10-1)
at Yukawa Institute for Theoretical Physics 
for stimulating discussion and presentations.
This work was supported in part by Monbukagaku-sho 
Grant-in-Aid for the Global COE programs, 
``The Next Generation of Physics, Spun from Universality 
and Emergence'' at Kyoto University and ``Quest for Fundamental Principles
in the Universe: from Particles to the Solar System and the Cosmos'' at
Nagoya University.
This work was also supported by JSPS Grant-in-Aid for Scientific Research (A) No.~18204024
and by Grant-in-Aid for Creative Scientific Research No.~19GS0219.
KT was supported by Grand-in-Aid for Scientific Research No.~21840028.
DY was supported by Grant-in-Aid for JSPS Fellows No.~20-1117. 
YS was supported by JSPS Postdoctoral Fellowships for Research Abroad.


\end{document}